\documentclass[usenatbib]{mn2e}
  \usepackage[T1]{fontenc}
  \usepackage{graphicx,color}
  \usepackage{subfigure}
  \usepackage{afterpage}
\graphicspath{{./FIG/}}

\title[Pulsar Dispersion Measure]{The effect of Pulse profile evolution on Pulsar Dispersion Measure}
\author[A.L. Ahuja et al.] 
{A. ~L. ~Ahuja,$^1$ D. Mitra,$^2$ and Y. Gupta$^2$ 
\\
$^1$ IUCAA, Ganeshkhind, Pune University, Pune, India \\
$^2$ National Center for Radio Astrophysics, TIFR, Pune University Campus, Pune 411007, India}
\date{Released 2007 Xxxxx XX}

\begin{document}
\maketitle

\pagerange{\pageref{firstpage}--\pageref{lastpage}} \pubyear{2007}
\def\LaTeX{L\kern-.36em\raise.3ex\hbox{a}\kern-.15em
    T\kern-.1667em\lower.7ex\hbox{E}\kern-.125emX}

\begin{abstract}
In an earlier paper \citep[][]{ahuja_etal_2005}, based on simultaneous 
multi-frequency observations with the Giant Metrewave Radio Telescope (GMRT),  
we reported the variation of pulsar dispersion measures (DMs) with frequency.  
A few different 
explanations are possible for such frequency dependence, and a possible 
candidate is the effect of pulse shape evolution on the DM estimation 
technique. In this paper we describe extensive simulations we have done 
to investigate the effect of pulse profile evolution on pulsar DM estimates. 
We find that it is only for asymmetric pulse shapes that the DM estimate 
is significantly affected due to profile evolution with frequency.
Using multi-frequency data sets from our earlier observations, we have carried 
out systematic analyses of PSR B0329+54 and PSR B1642$-$03.  Both these pulsars
have central core dominated emission which does not show significant 
asymmetric profile evolution with frequency.  Even so, we find that the 
estimated DM shows significant variation with frequency for these pulsars.
We also report results from new, simultaneous multi-frequency observations of 
PSR B1133+16 carried out using the GMRT in phased array mode.  This pulsar
has an asymmetric pulse profile with significant evolution with frequency.  
We show that in such a case, amplitude of the observed DM variations can be 
attributed to profile evolution with frequency.  
We suggest that genuine DM variations with frequency could arise due to 
propagation effects through the interstellar medium and/or the pulsar magnetosphere.
\end{abstract}


\section{Introduction }                 \label{sec:intro}
Radio signals from a pulsar undergo dispersion due to the ionized inter-stellar plasma.
This effect is characterised by the dispersion measure (DM) of the pulsar.  
In an earlier paper \citep[][hereafter {\it Paper I}]{ahuja_etal_2005}, we 
described a novel technique for estimation of pulsar DM using the simultaneous multi-frequency
observing capability of the GMRT.  This technique allows accurate estimates of
DM to be obtained at single epochs, without use of absolute timing information.
The accuracy of these single epoch DM estimates was shown to be $\sim$ 1 part in $10^4$
or better, for the bright pulsars.  
%This made for several interesting conclusions
%to be drawn from the study carried out in {\it Paper I}.

One of the interesting results reported in {\it Paper I} is
based on the 
%ability of the GMRT to observe  
fact that simultaneous dual-frequency observations, 
at more than one pair of radio frequency bands, 
were carried out 
within a short interval of time.  As a result, for some of the pulsars studied,
we could obtain DM estimates using different pairs of frequency bands, at the same
epoch.  The results from this showed small but significant differences in the DM
values from different frequency bands, for the pulsars thus studied.  This difference 
was of the same magnitude and sign at different epochs, signifying that it is a 
systematic effect rather than a random difference that changes with epoch.  There have 
been reports in literature about differences in pulsar DMs estimated from different 
parts of the radio spectrum \citep[e.g.][]{Shitov_etal,Hankins-91}.

A few different explanations are possible for frequency-dependent DM variations.  
A likely candidate is the effect of profile shape evolution with frequency  \citep[e.g.][]{PnW-92}, which
can lead to an effective shift of the phase of the pulse at different frequencies,
resulting in errors in estimating the dispersion delays between different frequency pairs.
On the other hand, physical origins for DM variations with frequency can be found both in 
the ISM and/or the pulsar magneto-spheric emission processes \citep[e.g.][]{Kardashev_etal}.

In this paper, we concentrate on a detailed study to assess the effect of pulse shape
evolution on DM estimates of pulsars. In section~\ref{sec:mul_obs} we discuss the data 
used for our study as well as discuss the new, simultaneous 4 frequency observations 
carried out using the GMRT.
To examine the role of the effects mentioned above, and to better understand the 
problem, numerical simulations can provide a powerful tool.  
The second part of this paper describes such simulations and discusses 
the conclusions obtained from them.

%%%%%%%%%%%%%%%%%%%%%%%%%%%%%%%%%%%%%%%%%%%%%%%%%%%
\section{Data and Observations} \label{sec:mul_obs}
%%%%%%%%%%%%%%%%%%%%%%%%%%%%%%%%%%%%%%%%%%%%%%%%%%%

To study frequency dependent DM variations, we focus on three bright pulsars
- PSR's B0329$+$54, B1642$-$03 and B1133$+$16 - each showing varying degree of 
profile complexity and pulse shape evolution with frequency.
Some useful parameters of these pulsars are given in Table \ref{tab:pul_param}.
Column 2 of Table \ref{tab:pul_param} contains the formal DM values given in
the old pulsar catalog \citep[][]{cat-93} and the new catalog 
\citep[][]{Hobb_etal_2004}, while Column 3 of the table displays the mean DM 
values from {\it Paper I} and Column 4 gives the pulsar period.  

For our study in this paper, we have used single epoch, 
simultaneous multi-frequency data for 
PSR's B0329$+$54 (at 610+320+243 MHz) and B1642$-$03 (at 610+325 and 325+243 MHz), 
based on observations reported in {\it Paper I}. 
Although the results for PSR B0329$+$54 reported in {\it Paper I} used only one 
dual-frequency observation at each epoch, at three epochs (January 8, January 22 
and May 14, 2001) this pulsar was observed at 3 frequency bands simultaneously: 
610+320+243 MHz.  Using the data from these epochs, we have estimated the DM
values for the different frequency pairs.  The results for the January 22, 2001 
observations give DM values of 26.78610(4) pc cm$^{-3}$ for the 610+320 MHz 
frequency pair, 26.77947(3) pc cm$^{-3}$ for the 610+243 MHz frequency pair 
and 26.77256(5) pc cm$^{-3}$ for the 230+243 MHz frequency pair.
For PSR B1642$-$03, using the data from August 18, 2001, we obtain DM values of
35.75809(7) pc cm$^{-3}$ for the 610+325 MHz frequency pair and 35.72262(7) pc cm$^{-3}$ 
for the 325+243 MHz frequency pair.
For PSR B1133$+$16, observations reported in {\it Paper I} had only one frequency 
pair at each epoch.  Hence, to obtain simultaneous multi-frequency data, we made 
new observations with the GMRT using an improved technique, as described below.

In {\it Paper I} we described the basic scheme for using the
GMRT in a simultaneous dual-frequency pulsar observing mode, as well as the
steps involved in processing the data for obtaining accurate estimates of pulsar
DMs.  As explained there, the crucial aspect of the scheme is that, in order to
eliminate all instrumental delays, baseband signals from two different radio
frequency bands (coming from two sub-arrays of the GMRT) are added together in
one multi-channel pulsar receiver. The dispersion delay of the pulsar signal
between the two radio frequency bands is then used for discriminating between
the pulses at the two frequency bands.  One drawback of this scheme was that
the pulsar signals from the two frequency bands could overlap each other, depending
on the nature of the dispersion delay track across the 16 MHz of baseband bandwidth
of each of the two frequency bands (see {\it Paper I} for details).
This makes it difficult to apply the scheme easily for simultaneous observations
using 3 or 4 frequency bands.

To overcome the above limitation, we now use a modified scheme wherein the digital
sub-array combiner which does the addition of the multi-channel baseband
data from different antennas, is programmed to blank the data for a selected set
of frequency channels for antennas from a given sub-array.  This is akin to setting
non-overlapping filter banks for each sub-array and allows the 16 MHz bandwidth to
be divided between the different radio frequency bands of observations, without any
overlap between signals from the different bands, while still preserving the time
alignment of the data from the different frequency bands.  The sub-band for each
frequency band can be conveniently placed anywhere in the 16 MHz band, by simply
changing the channel mask settings for each antenna of the corresponding sub-array.
Using this method, the number of frequency bands for which simultaneous multi-frequency
observations can be done is easily extended -- it is limited only by the loss of S/N
due to the reduction of the effective bandwidth from each frequency band.  This loss
is compensated to some extent by the fact that in this new scheme, the addition of
the data from the different sub-arrays can be carried out in the voltage domain
(instead of incoherent addition in the older scheme), allowing the use of phased
array mode of operations for each sub-array.

Using the new scheme, we carried out simultaneous observations at four frequency
bands of the GMRT -- 235, 325, 610 and 1280 MHz -- for pulsar B1133$+$16, on February 3,
2004. The pulsar was found to be weak at 1280 MHz and did not have sufficient signal
to noise for reliable DM estimation using this frequency. Hence we restricted our 
study to 3 frequencies: 235, 325 and 610 MHz. 
The total bandwidth for each observing frequency band was 4 MHz. 
The observation was carried out with a sampling interval of 0.256 ms, which is a 
factor of 2 better than the 0.516 ms used for the observations reported in {\it Paper I}.
The results of the observations give DM values of 4.8098(2) pc cm$^{-3}$ for the 
610+325 MHz frequency pair, 4.8311(1) pc cm$^{-3}$ for the 610+243 MHz frequency pair 
and 4.8290(2) pc cm$^{-3}$ for the 325+243 MHz frequency pair.

%%%%%%%%%%%%%%%%%%%%%%%%%%%
\begin{table*}
\begin{minipage}{170mm}
\caption[Sample of pulsars]{ Important parameters of the sample of pulsars.}
\label{tab:pul_param}
\begin{tabular}{|c|r|r|r|}
\hline
Pulsar  & Catalogue DM  & Measured & Period \\
name    & (Old/New)     & mean DM\footnote{see {\it Paper I}} & \\
        & ($\rm pc ~cm^{-3}$) & ($\rm pc ~cm^{-3}$) & ($\rm sec$) \\ \hline
B0329$+$54 & 26.776/26.833 & 26.77870(3)  &  0.7145 \\
B1133$+$16 &  4.8471/4.864 & 4.8288 (6)   &  1.1877 \\
B1642$-$03 & 35.665/35.727 & 35.75760(14)\footnote{from frequency pair 610$+$325 MHz} & 0.3877 \\
B1642$-$03 & 35.665/35.727 & 35.72270(7)\footnote{from frequency pair 325$+$243 MHz} &  0.3877 \\
\hline
\end{tabular}
\medskip
\end{minipage}
\end{table*}
%%%%%%%%%%%%%%%%%%%%%%%%%%%

%%%%%%%%%%%%%%%%%%%%%%% 
\section{Factors affecting DM estimates} \label{sec:factors}
%%%%%%%%%%%%%%%%%%%%%%% 
 
In order to understand the intriguing result of DM varying with frequency, we 
need to assess the primary factors affecting DM estimates.  Let us briefly recall 
the DM estimation technique which has been described in detail in {\it Paper I}, 
and then look into the factors which can produce the observed DM change.
Basically, the DM is estimated from the delay in the pulse arrival time $\Delta t_{m}$ 
at two frequencies $f_1$ and $f_2$, and is given by
\begin{equation}
DM = \frac{ \Delta t_{m}}{K \left( \frac{1}{f_{1}^{2}} - \frac{1}{f_{2}^{2}} \right)}
\label{eq1}
\end{equation}
where the constant $K = 1/( 2.410331 \times 10^{-5})$.  
All the DM results reported in this paper are based on using the average profile (AP) 
method described in {\it Paper I}, where the DM between two simultaneously observed 
frequency bands is obtained based on estimation of the arrival time delay between the 
average profiles at the two different frequency bands. As per equation 7 in {\it Paper I}, 
the total delay $\Delta t_{m}$ can be written as a combination of 3 terms : 
$\Delta t_{m} = \Delta t_p + \Delta t_i + \Delta t_f$, where $\Delta t_p$ is the integral 
number of pulsar periods in the total delay, $\Delta t_i$ is the integral number of time 
samples of delay within a pulsar period, and $\Delta t_f$ is the fractional sample time 
component of the delay. 
One of the main difficulties in estimating the time delays is the lack of knowledge of an 
appropriate fiducial or reference point in the pulse profiles at different frequencies. 
In the absence of this, the procedure adopted to find the delay is by cross-correlating 
the pulse profiles at the two frequency bands. In the AP method, $\Delta t_p$ is calculated
based on the knowledge of the frequencies for the two bands, the catalog DM value and the 
pulsar period. The value of $\Delta t_i$ is obtained by finding the lag at which the peak 
of the cross-correlation occurs. The profile at the lower frequency is then rotated left
circularly by this amount to align it with the higher frequency pulse profile. Finally, 
$\Delta t_f$ is estimated (refer to Equations 5 to 15 of {\it Paper I} and the discussion 
therein for details) from the cross-spectrum, $CS(\nu)$, of these aligned profiles. The 
phase of the cross-spectrum (hereafter CS) can be written as, 
\begin{equation}
\phi_{CS}\left(\nu\right) ~~=~~ \phi_{2i} - \phi_{1i} + 2\pi\nu\Delta t_f
\label{eq2}
\end{equation}
where 
$\phi_{1i}$ and $\phi_{2i}$ are the intrinsic phase functions of the two pulse profiles. 
Thus, for $\phi_{1i} = \phi_{2i}$ i.e. when the pulse profiles at the two frequencies have 
the same shape, $\Delta t_f =  \frac{\Delta\phi_{CS}}{2\pi\Delta\nu}$.  Here, 
$\Delta\phi_{CS}$ is the phase change in the cross-spectrum, across an interval 
of $\Delta\nu$.

In the above analysis procedure, the estimated DM can show a frequency dependence if the
profile changes significantly with frequency.  
The effects of these DM changes will be reflected in modified values for $\Delta t_i$ and/or  
$\Delta t_f$.  Now, it is well known that pulse widths at lower frequencies are more than 
those at higher frequencies -- a phenomenon referred to as `radius-to-frequency mapping' (RFM) 
-- see \cite{Mit_Ran_2002} for a recent review.  Furthermore, for several pulsars, pulse 
shapes at various frequencies can differ significantly \citep[see][]{Ran_1983}. 
For complex pulse shapes with multiple emission components, the relative strengths of the 
components, and even the total number of components, can change with frequency. Such pulse 
shape changes can cause the peak of the amplitude of the cross-correlation function to shift 
resulting in change in the value of $\Delta t_i$.  Further, pulse shape variations with 
frequency can lead to intrinsic pulse phases, $\phi_{1i}$ and $\phi_{2i}$, 
being different, causing $\Delta t_f$ to change (see Equation \ref{eq2}). 
Thus, it is important to be able to estimate the magnitude of DM changes due to profile 
evolution with frequency, before looking for other possible origins for the observed DM 
changes with frequency.  In the next section we describe and present the results of 
simulations that we have carried out to disentangle the effect of pulse shape evolution on 
DM estimates. 

Before we close this section, it is important to dwell on the issue of the fiducial point 
in the pulse with respect to which DMs are found. In this aspect, the scheme we have 
employed here to find the DM is slightly different from most of the previously used 
techniques. The difference lies in the choice of the fiducial point of the profile. Ideally, 
to find accurate DMs, the time of arrival estimate of the pulse at different frequencies 
needs an identifiable feature in the pulse waveform which corresponds to the same rotational 
phase of the pulsar at each observing frequency.
For example, for double component profiles, a number of authors \citep[e.g.][]{Cra_1970, PnW-92} 
have used the midpoint between the outer peaks as the fiducial point; and for single component
profiles, the peak of the pulsar profile is used as the fiducial point. 
Based on simultaneous observations of pulse micro-structure, \cite{Bor_1983} has shown
that the micro-structure in a single pulse is not affected by dipolar field line
spreading and can be used as accurate fiducial marker for measurement of DMs.  Further, for 
the double component profile pulsar PSR B1133+16, he found that the micro-structure 
based fiducial marker is in good agreement with the midpoint between conal peaks.
In contrast to the above, the CS method that we have used, by its nature, tries to 
align the profile with respect to the centroid of the pulse. For highly asymmetric
profiles the centre and the centroid can be different, thus leading to different 
DM estimates. The simulations discussed in the next section try to resolve some of 
these issues.
 
%
%%%%%%%%%%%%%%%%%%%%%%%%%%
\section{Simulations} \label{sec:simul}
%%%%%%%%%%%%%%%%%%%%%%%%%%
%
With the objective of quantifying the effect of pulse shape variations on
DM estimates, we have carried out extensive simulations.
The basic paradigm of these simulations is to construct various complex pulse shapes
at different frequency bands with a user supplied DM value, pass them through
our analysis technique to obtain an output DM value, and compare it with the input DM.
Under these controlled situations it is possible to try and pin down the 
cause for DM changes, at various stages of the analysis.

We have simulated pulse profiles with the observing parameters tuned for the 
GMRT observations. The following input parameters were used for the simulations: 
(1) pre-decided pulsar period, $P$; 
(2) the following information about the frequency bands:
(a) frequency band values matching with the real observations, 
(b) band-width, used in the real observations, 
(c) number of channels (typically 256) across the band-width, as used in the real observations, 
(d) side band corresponding to lower side band (USB) or upper side band (LSB), as used in the real observations; 
(3) time sample bin size, $T$; 

At a given frequency, pulse profiles were generated with integral 
number of time sample bins per period, $N ~=$ nearest integer$(P/T)$,
and hence the effective time sampling was $T_{eff} ~=~ P/N$.

It is well known that profiles of several pulsars are composed of a number 
of emission components \citep[see]{Ran_1993}. 
Detailed pulse shape phenomenological studies have revealed that, in general,  
pulsar emission can be interpreted in the form of a central core emission with 
nested cones around it \citep[see][]{Mit_Des_1999}. 
Depending on how the line-of-sight cuts through the emission cone,
a given pulse profile might contain an odd number of components where emission 
from both the cone and core is present, or an even number of components
when only the conal emission is seen. Also, the conal emission
pairs can have varying intensity and sub-pulse widths.
The centre of the core emission is often used as a fiducial point to mark 
the centre of the profile.  Due to aberration and retardation effects, the 
mid-point of the conal components is expected to lead the centre of the core 
emission \citep[e.g.][]{Gan_Gup_2001}.

In our simulations, we have used the aforementioned pulsar properties to generate 
the pulse profiles.  Assuming that the emission from the core and the cones have a 
Gaussian intensity distribution \citep{kramer_etal_1994}, we have generated pulse 
profiles where the profile is composed of a number of components with each component 
being a Gaussian function.  To generate a profile with a given signal to noise ratio,
Gaussian random white noise was added to the signal.  Based on the above considerations, 
the signal at the $i^{th}$ bin of the profile is calculated as,
\begin{equation}	
y(i) = \sum_{j=1}^n [A(j) ~ \times~ e^{-(x(i) - \langle x(j)\rangle)^2/(2\sigma(j)^2)}] \end{equation}
Here, $y(i)$ is the $i^{th}$ bin signal. The index $j$ refers to the number of peaks 
in a profile, $\langle x(j)\rangle $ is position of the $j^{th}$ peak (centre of the Gaussian), 
$A(j)$ is the amplitude of the $j^{th}$ peak, and $\sigma (j)$ is the RMS of the $j^{th}$ peak 
function.

To produce profiles at multiple frequencies, we have used the conventional wisdom 
about pulse width evolution properties, where it is generally seen that the overall 
pulse width scales with frequency $\nu$ as $\nu^{-\alpha}$, where $\alpha$ is a 
positive spectral index, with a typical value $\sim 0.2$ \citep[see][]{Mit_Ran_2002}. 
To get the delayed pulse corresponding to 
an input DM, a fixed fiducial point in the profile needs to be chosen. For profiles 
with an odd number of components, we have chosen the peak of the central core component 
as the fiducial point.  For profiles with even number of components, the fiducial point 
was chosen to be the mid-point of the outer conal component locations.
The data generated at the two frequencies were passed through the DM estimation code,
to obtain our DM results. 

%%%%%%%%%%%%%%%%%%%%%%%%%%%%%%%%%%%%%%%%%%%%%%%%
\subsection{Simulation results} \label{subsec:sim_res}
%%%%%%%%%%%%%%%%%%%%%%%%%%%%%%%%%

The simulations were carried out in different steps, starting from a single peak 
profile without any frequency evolution, and extending to multi-component profiles
with significant frequency evolution.  Results from the simulation experiments are 
summarised in Table \ref{tab:simul_res}.

%\newpage
%%%%%%%%%%%Table 1%%%%%%%%%%%%%%%%%%%%%%%%%%%%%%%%%
\begin{table*}
\begin{center}
\caption[DM results]{DM results from simulations.}
\label{tab:simul_res}
\begin{tabular}{|c|c|c|c|c|c|c|c|} \hline
{\bf 1$^1$}&{\bf 2}&{\bf 3}&{\bf 4}&{\bf 5}&{\bf 6}&{\bf 7}&{\bf 8} \\ 
\hline

1 & Single peak, no evolution & 26.776 & 0.00 & 2.65, 0.055, 0.0 & 1302, 0.49123, 0.51594, 0.67201 & 26.77589(13) & $-$0.837 \\
%AVGPROF/SNGLPK/BEFOREMOD
\hline
2 & Single peak with evolution & 26.776 & 0.25 & 2.65, 0.055, 0.0 & 1302, 0.49227, 0.51594, 0.67201 & 26.77591(14) & $-$0.648 \\
%AVGPROF/SNGLPK/BEFOREMOD
\hline
3 & Symmetric, 2 peaks, & 26.776 & 0.00 & 2.65, 0.055, $-$5.0 & 1303, -0.4957, 0.5159, 0.6720 & 26.7762(1) & 1.67 \\
%AVGPROF/TWOPK/SYM/AFTERMOD/03012006/Without_evol
  & no evolution &              &     & 2.65, 0.055, 5.0 &           &          &           \\
\hline
4 & Symmetric 2 peaks, & 26.776 & 0.25 & 2.65, 0.055, $-$5.0 & 1303, -0.4940, 0.5159, 0.6720 & 26.7762(1) & 1.67 \\
%AVGPROF/TWOPK/SYM/AFTERMOD/03012006/With_evol
  & with evolution &        &      & 2.65, 0.055, 5.0    &          &         &            \\
\hline
5 & Asymmetric 2 peaks & 26.776 & 0.00 & 1.00, 0.035, $-$5.0 & 1303, -0.47532, 0.51594, 0.67203 & 26.77666(34) & 1.58  \\
%AVGPROF/TWOPK/ASYM/AFTERMOD/03012006/Without_evol
  & no evolution &              &     & 2.00, 0.055, 5.0  &           &          &             \\
\hline
6 & Asymmetric 2 peaks, & 26.776 & 0.25 & 1.00, 0.035, $-$5.0  & 1303, 0.3775, 0.5159, 0.6725&26.7941(4) & 48.01 \\
%AVGPROF/TWOPK/ASYM/AFTERMOD/03012006/With_evol/InDM26776
  & with evolution &        &      & 2.00, 0.055, 5.0   &         &        &           \\
\hline
7 & 3 peak symmetric & 26.776 & 0.20 & 2.300, 0.186, $-$10.541 & 1302,  0.4917,  0.5159, 0.6720 & 26.7759(1) & $-$1.06 \\
%AVGPROF/B1133/18_AUG_2001_FIT/Thr_Pk/SYM
  &  &          &     & 12.146, 0.040, 0.0 &           &          &               \\
  &  &          &     & 2.300, 0.186, $-$10.541 &           &          &             \\
\hline
8& 3 peak asymmetric& 26.776& 0.20& 2.300, 0.186, $-$11.249& 1302,  0.4061, 0.5159, 0.6720  &26.7741(1) & $-$18.66 \\
%AVGPROF/B1133/18_AUG_2001_FIT/Thr_Pk/ASYM/ASYM_SEP
  & separation     &            &     & 12.146, 0.040, 0.0     &           &                 &      \\
  &                &            &     & 2.300, 0.186, $-$10.541 &           &                 &     \\
\hline
9& 3 peak asymmetric & 26.776& 0.20& 2.300, 0.186, $-$11.249& 1302, -0.2586, 0.5159, 0.6716 &26.7605(1) & $-$150.67 \\
%AVGPROF/B1133/18_AUG_2001_FIT/Thr_Pk/ASYM/ASYM_SEP_SIG
  & separation     &    &      & 12.146, 0.040, 0.0&              &           &         \\
  & and sigma      &            &     & 4.019, 0.186, $-$10.541 &           &                 &     \\
\hline
10& 3 peak asymmetric separation, & 26.776 & 0.20 & 2.300, 0.186, $-$11.24941 & 1300, 0.0582, 0.5159, 0.6708  &26.7259(1) &$-$415.35 \\
%AVGPROF/B1133/18_AUG_2001_FIT/Thr_Pk/ASYM/ASYM_SEP_SIG_AMP/05052005
  & sigma and      &            &       & 12.146, 0.040, 0.0 &                  &           &      \\
  & amplitude      &            &       & 4.019, 0.070, $-$10.541 &                          &     & \\
\hline
\end{tabular}
\end{center}
{$^{1}$For detail description of the contents in all the columns, 
see paragraph 2 in Section \ref{subsec:sim_res} }  \hfill{ }
\vspace{5mm}
\end{table*}
%%%%%%%%%%%%%%%%%%%%%%%%%%%%%%%%%%%%%%%%%%%%

The information in the different columns (1 to 8) of this table, is as follows:
\noindent
%\noindent {\underline{\bf Column:} \ {\underline{\bf Description}} \\

\noindent{\bf 1.} \hspace{10mm}	Experiment number; \\
{\bf 2.} \hspace{10mm}	About the type of profile generated in the experiment; \\
{\bf 3.} \hspace{10mm}	Input DM (in $\rm pc ~cm^{-3}$); \\
{\bf 4.} \hspace{10mm}	Pulse profile evolution index; \\
{\bf 5.} \hspace{10mm}	Parameters of each component (a Gaussian function): sigma, $\sigma$ 
(in ms), amplitude, separation (in ms) of peak from reference point; \\
{\bf 6.} \hspace{10mm}	Total integral number of time samples of delay, fractional sample 
time component of delay, effective time sample bin size (in ms) and 
total measured time delay (in s), as estimated from our DM analysis; \\
{\bf 7.} \hspace{10mm}	Output DM  (in $\rm pc ~cm^{-3}$) from the experiment, 
with $1\sigma$ error bar; \\
{\bf 8.} \hspace{10mm}	Deviation of the output DM from the input DM, in units of $\sigma$. \\
%%%%%%%%%%%%%%%%%%%%%%%%%%%%%%%%%%%%%%%%%%%%%%%%%%%%%%%%
%

%%%%%%%%%%%%%%%%%%%%%%%%%%%%%%%%%%%%%%%%%%%%%
\begin{figure*}
\begin{center}
\includegraphics[angle=-90, width=0.6\textwidth]{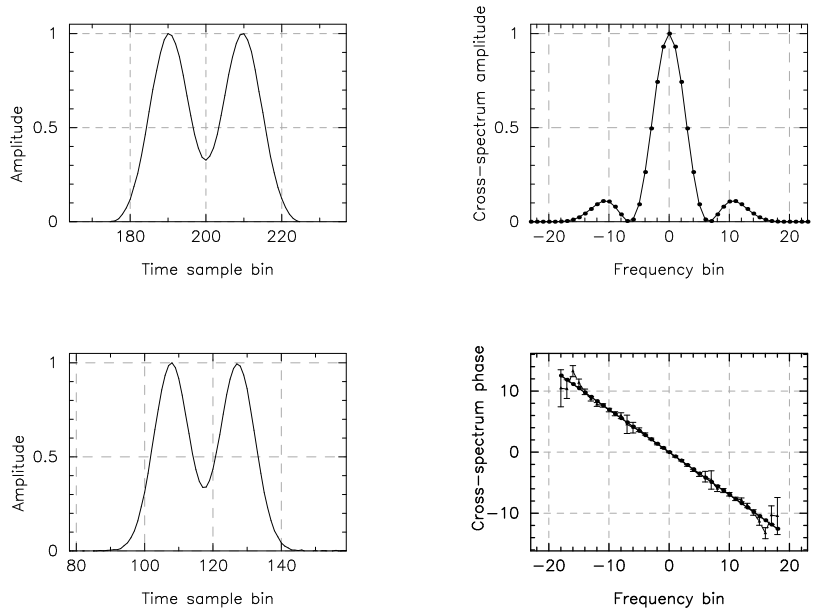}
\caption[2 peak symm no evol simul]
%{Simulation with two peaks, symmetric profile without 
%evolution. 
{Two peaks, symmetric, without evolution, simulated 
pulse profiles at the two frequencies $325$ 
(left lower panel) and $610 ~\rm MHz$ (left upper panel), 
and the corresponding CS 
amplitudes (right upper panel) and phase with best fit line (right lower panel).}
\label{fig:2psn}
\end{center}
\end{figure*}
%%%%%%%%%%%%%%%%%%%%%%%%%%%%%%%%%%%%%%%%%%%%%
%%%%%%%%%%%%%%%%%%%%%%%%%%%%%%%%%%%%%%%%%%%%%
\begin{figure*}
\begin{center}
\includegraphics[angle=-90, width=0.6\textwidth]{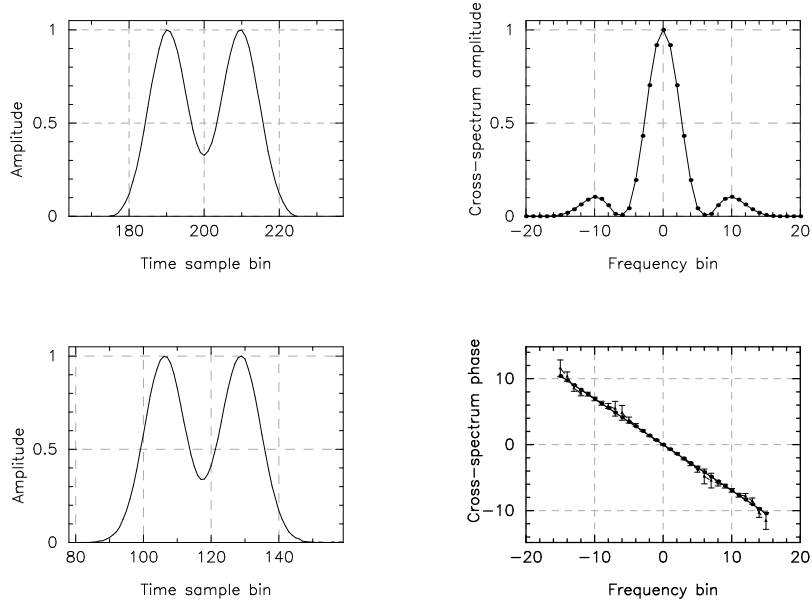}
\caption[2 peak symm no evol simul]
%{Simulation with two peaks, symmetric profile with 
%evolution. Pulse profile at the two frequencies $325$ 
%(left lower panel) and $610 ~\rm MHz$ (left upper panel), and corresponding CS 
%amplitudes (right upper panel) and phase with best fit line (right lower panel).}
{The left lower and upper panels represent the two peaks, symmetric, simulated pulse profiles 
at the two frequencies 325 and 610 MHz, respectively. 
The corresponding CS amplitude and phase function with best fit are shown
in the right upper and lower panels, respectively.}
\label{fig:2psw}
\end{center}
\end{figure*}
%%%%%%%%%%%%%%%%%%%%%%%%%%%%%%%%%%%%%%%%%%%%%
%
%%%%%%%%%%%%%%%%%%%%%%%%%%%%%%%%%%%%%%%%%%%%%
\begin{figure*}
\begin{center}
\includegraphics[angle=-90, width=0.6\textwidth]{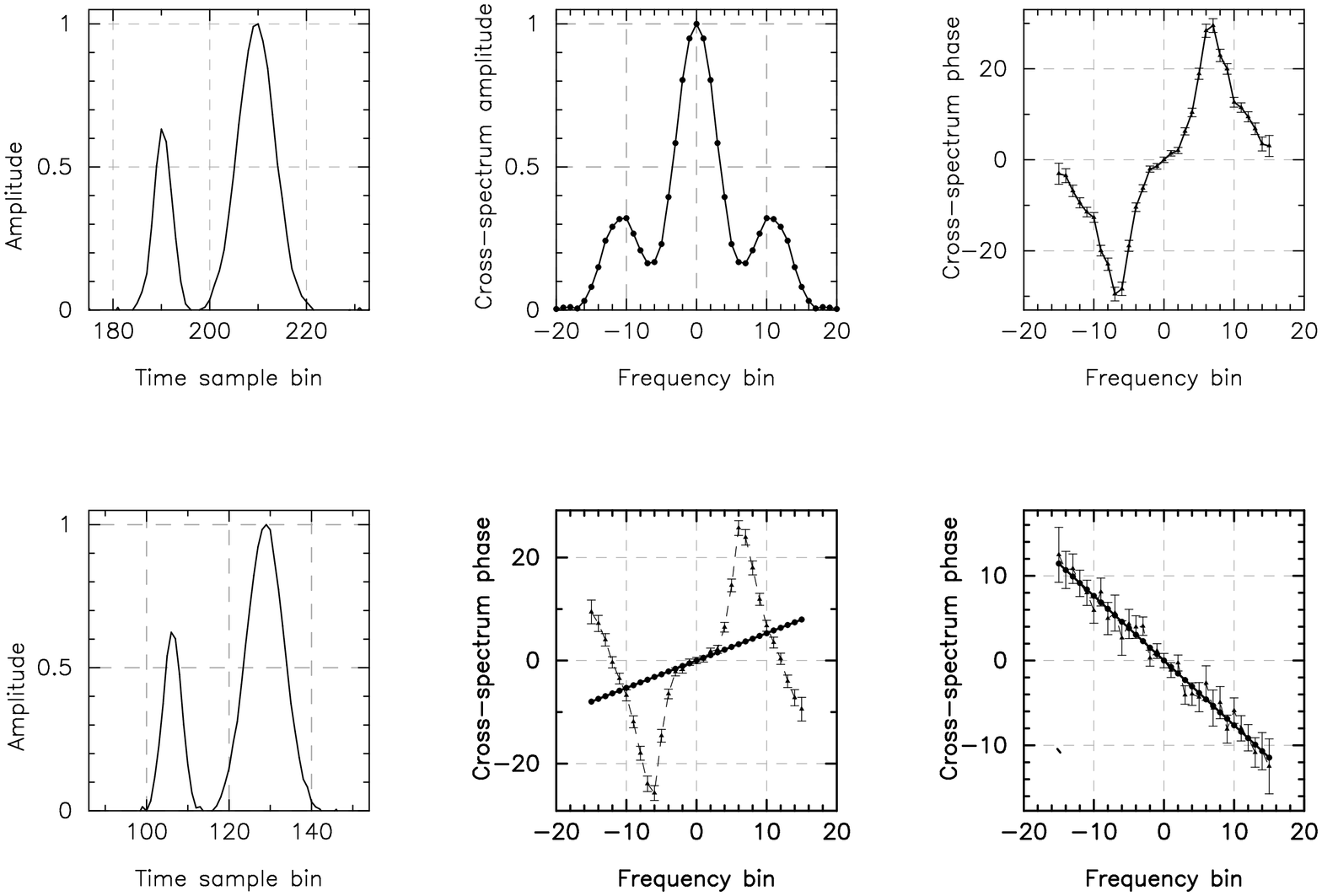}
\caption[Simulation: 2 peaks, symmetric, no evolution]{The figure shows two peak, asymmetric,
with evolution, simulated pulse profiles at two frequency bands $325$
(left lower panel) and $610 ~\rm MHz$ (left upper panel) respectively. The corresponding CS
amplitude (middle upper panel), phase function with best fit line (middle lower panel),
template phase function (right upper panel) and template subtracted phase function with
best fit line (right lower panel) are shown in the figure.}
\label{fig:2paw}
\end{center}
\end{figure*}
%%%%%%%%%%%%%%%%%%%%%%%%%%%%%%%%%%%%%%%%%%%%% 
%

All the simulations were carried out for a pulsar with period of $714.5801473327570 ~\rm ms$
and DM of $26.776 ~\rm pc~cm^{-3}$ respectively\footnote{The results obtained from our 
simulations are independent of these input values.
To compare the simulation results with a realistic case,
we have used these values which correspond to one of the sample pulsars PSR B0329$+$54.}.
The RMS of the additive noise was kept relatively low (0.015) for the initial experiments 
(simulation numbers 1 to 4); but in later experiments (simulation numbers 5-10), to see 
the effect of lower SNR, the input RMS value was increased to 0.05.  Profiles were 
generated at the frequency bands of 610 and 325 MHz. In the analysis, for fitting the 
phase of the CS with a  straight line to 
estimate the fractional sample time delay, the number of data points used on each side 
of the $0$ frequency channel were $\sim 20$.

The simulations presented in Table \ref{tab:simul_res} were carried out systematically, 
by adding complexities in each successive simulation. Here we discuss the results of
3 simulations in detail, namely simulation numbers 3, 4 and 6, which illustrates the 
effect of pulse profile evolution on DM estimates. The results of simulation 3 are 
presented in figure~\ref{fig:2psn}. In this case we generated pulse profiles with 
two symmetrical peaks and the profiles are not evolving in widths with frequency. 
As is seen in the figure, the corresponding phase function of the CS shows a linear 
gradient (bottom right panel) as expected, which can be well fit with a straight
line to obtain the fractional delay $\Delta t_f$ with significant accuracy. This is 
primarily because of the cancellation of the phase function of the Fourier transformed 
profiles, in the CS phase i.e. $\phi_{1i} = \phi_{2i}$ in equation~\ref{eq2}. 
Figure~\ref{fig:2psw} shows the case of simulation 4 where we consider two component,
symmetric pulse profiles.  The widths of the components and the overall profile 
evolve with frequency with a spectral index $\alpha$ of 0.25, with the widths increasing 
at lower frequencies.  Even in this case, the cancellation of the phase functions leads 
to a linear phase of the CS, allowing for accurate DM estimation. A similar argument 
holds for simulation numbers 1 through 5 and simulation number 7, as can be seen in 
Table \ref{tab:simul_res}.  For all of these cases, the output DM is in good agreement 
with the input DM.

In simulation 6 two component profiles were generated with a significant degree of 
asymmetry.  The profile and component widths were evolved with frequency with a spectral 
index of 0.25.  Under these circumstance the intrinsic phase function of the profile at
each frequency is different and hence the phase function of the CS acquires a complicated 
phase superposed on the linear gradient due to the fractional delay. This is illustrated 
in figure~\ref{fig:2paw} where the complexity in the phase function is apparent (bottom 
middle panel). Attempts to obtain a DM by fitting a straight line to this phase function 
yield a significantly different value of DM (48$\sigma$ deviation), as seen in column 8 
of Table \ref{tab:simul_res}.  Results of simulation numbers 8 through 10, where the 
profiles have 3 components and are asymmetric, given in Table~\ref{tab:simul_res}), show 
DM inconsistencies due to similar reasons.

However, if the phase contributions due to the individual pulse profiles are known, it 
should be possible to subtract these from the CS phase and recover the linear gradient. 
In the simulations we can do this by creating zero DM profiles using the same fiducial 
point that has been used to simulate the finite DM profiles.  The CS phase for the zero 
DM profiles is then a measure of the phase function due to the intrinsic pulse shapes.  
This is shown in the top right corner of figure~\ref{fig:2paw}. This phase function, 
when subtracted from the CS phase of the finite DM profiles, gives back the linear 
gradient -- as shown in the bottom right corner of figure~\ref{fig:2paw}.  The DM 
estimated using this gradient is $26.7752(5) ~\rm pc~cm^{-3}$, 
and is within $2\sigma$ of the input DM. This is a significant improvement from the 
$48\sigma$ discrepancy, found before correction.

Note that application of the zero DM template phase subtraction technique to real data, 
needs an assumption for the fiducial point. However, although the profiles are aligned 
with respect to the fiducial point, the CS method actually finds the DMs by aligning the 
centroid of the profiles.  Thus, for cases where the difference between the fiducial point 
and the centroid of the profile varies considerably with frequency, the CS method would 
yield a finite DM even for zero DM aligned profiles. Nonetheless, the template subtracted 
DM will eventually be based on the choice of the fiducial markers. In principle, for the 
correct choice of the fiducial marker, the phase subtracted template should recover the 
linear phase gradient which is solely due to the time delay effect.

In summary, our simulations show that pulse shape evolution, which results in significant 
differences in the intrinsic phase functions, can affect DM estimates in pulsars. 

%%%%%%%%%%%%%%%%%%%%%%%%%%%%%%%%%%%%%%%%%%%%%%%%%%%%
\begin{table*}
\begin{center}
\caption[DM results for PSR B1642$-$03]{DM results for PSR B1642$-$03 observed at two pairs of frequency bands 
610$+$325 and 325$+$243 MHz, at one of the epoch, August 18, 2001.}
\label{tab:b1642}
%\begin{tabular}{|c|c|c|c|c|} \hline
%Frequency  & Real data & $\Delta(DM)$ & Simulation & $\Delta(DM)$ \\
%Combination & Output DM ($\sigma_{DM}$)&  & Output DM ($\sigma_{DM}$) & \\
\begin{tabular}{|c|c|l|c|l|c|} \hline
Frequency  & DM/bin 		& DM ($\sigma$), before & $\Delta DM$ ($\Delta$bin)& DM ($\sigma$), after & $\Delta DM$ ($\Delta$bin) \\
Combination &  	& template subtraction &  & template subtraction & \\
MHz & $\rm pc~cm^{-3}/bin$	& $\rm pc~cm^{-3}$($\rm pc~cm^{-3}$)  &  $\rm pc~cm^{-3} (\#bin)$ & $\rm pc~cm^{-3}$  & $\rm pc~cm^{-3} (\#bin)$ \\
\hline
610$+$325 & 0.02058 & 35.75809(7) &    0.00000 (   0.000) & 35.75308(9) &    0.000 (   0.000)  \\
325$+$243 & 0.01493 & 35.72262(7) & $-$0.03547 ($-$2.377) & 35.72125(9) & $-$0.032 ($-$2.133)  \\
\hline
\end{tabular}
\end{center}
\end{table*}
%%%%%%%%%%%%%%%%%%%%%%%%%%%%%%%%%%%%%%%%%%%%%%%%%%%%%%%%%%%
%%%%%%%%%%%%%%%%%%%%%%%%%%%%%%%%%%%%%%%%%%%%%%%%%%%%%%%%%%%
\begin{table*}
\begin{center}
\caption[DM results for PSR B0329$+$54]{DM results for PSR B0329$+$54 observed simultaneously at three frequency bands
243, 320 and 610 MHz, on January 22, 2001}
\label{tab:b0329}
\begin{tabular}{|c|c|l|c|l|c|} \hline
Frequency  & DM/bin 		& DM ($\sigma$), before & $\Delta DM$ ($\Delta$bin)& DM ($\sigma$), after & $\Delta DM$ ($\Delta$bin) \\
Combination &  	& template subtraction &  & template subtraction & \\
MHz & $\rm pc~cm^{-3}/bin$	& $\rm pc~cm^{-3}$($\rm pc~cm^{-3}$)  &  $\rm pc~cm^{-3} (\#bin)$ & $\rm pc~cm^{-3}$  & $\rm pc~cm^{-3} (\#bin)$ \\
\hline
%1280$+$610 & 26.843(1)   &  0.00000 & 26.786(1)  & 0.000 \\
%1280$+$325 & 26.81421(7) & $-$0.02879 & 26.7799(3) & $-$0.0061 \\
%1280$+$227 & 26.8012(2)  & $-$0.0418  & 26.78003(13) & $-$0.00597 \\
%610$+$325 & 26.80407(4)  & $-$0.03893 & 26.77753(15) & $-$0.00847 \\
%610$+$227 & 26.79525(7)  & $-$0.04775 & 26.77843(7)  & $-$0.00757 \\
%325$+$227 & 26.78823(5)  & $-$0.05477 & 26.77844(5)  & $-$0.00756 \\
610$+$320 & 0.01724		& 26.78610(4)  &    0.00000 (   0.000) & 26.78404(6)  &    0.00000 (00.000) \\
610$+$243 & 0.00865		& 26.77947(3)  & $-$0.00663 ($-$0.766) & 26.77784(4)  & $-$0.00620 ($-$0.717) \\
320$+$243 & 0.01734		& 26.77256(5)  & $-$0.01360 ($-$0.784) & 26.77161(7)  & $-$0.01243 ($-$0.717) \\
\hline
\end{tabular}
\end{center}
\end{table*}
%%%%%%%%%%%%%%%%%%%%%%%%%%%%%%%%%%%%%%%%%%%%%%%%%%%%%%%%%%%
%
%%%%%%%%%%%%%%%%%%%%%%%%%%%%%%%%%%%%%%%%%%%%%%%%%%%%%%%%%%%
\begin{table*}
\begin{center}
\caption[DM results for PSR B1133$+$16]{DM results for PSR B1133$+$16 observed simultaneously at four frequency bands 
227, 325, 610 and 1280 MHz, on February 3, 2004. Data analysed at 3 frequency bands.}
\label{tab:b1133}
%\begin{tabular}{|c|l|l|c|c|} \hline
%Frequency  & Real data & $\Delta(DM)$ & Simulation & $\Delta(DM)$ \\
%Combination & Output DM ($\sigma_{DM}$)& & Output DM ($\sigma_{DM}$) & \\
\begin{tabular}{|c|c|l|c|l|c|} \hline
Frequency  & DM/bin 		& DM ($\sigma$), before & $\Delta DM$ ($\Delta$bin)& DM ($\sigma$), after & $\Delta DM$ ($\Delta$bin) \\
Combination &  	& template subtraction &  & template subtraction & \\
MHz & $\rm pc~cm^{-3}/bin$	& $\rm pc~cm^{-3}$($\rm pc~cm^{-3}$)  &  $\rm pc~cm^{-3} (\#bin)$ & $\rm pc~cm^{-3}$  & $\rm pc~cm^{-3} (\#bin)$ \\
\hline
610$+$325 & 0.0101 & 4.8098(2) & 0.0000 (0.000) & 4.8547(3) & 0.0000 (   0.000) \\
610$+$227 & 0.0043 & 4.8311(1) & 0.0213 (4.958) & 4.8521(2) & 0.0026 ($-$0.605) \\
325$+$227 & 0.0074 & 4.8290(2) & 0.0192 (2.597) & 4.8501(3) & 0.0046 ($-$0.619) \\
\hline
\end{tabular}
\end{center}
\end{table*}
%%%%%%%%%%%%%%%%%%%%%%%%%%%%%%%%%%%%%%%%%%%%%%%%%%%%%%%%%%%

%%%%%%%%%%%%%%%%%%%%%%%%%%%%%%%%%%%%%%%%%%%%%
\begin{figure*}
\begin{center}
\includegraphics[angle=-90, width=0.6\textwidth]{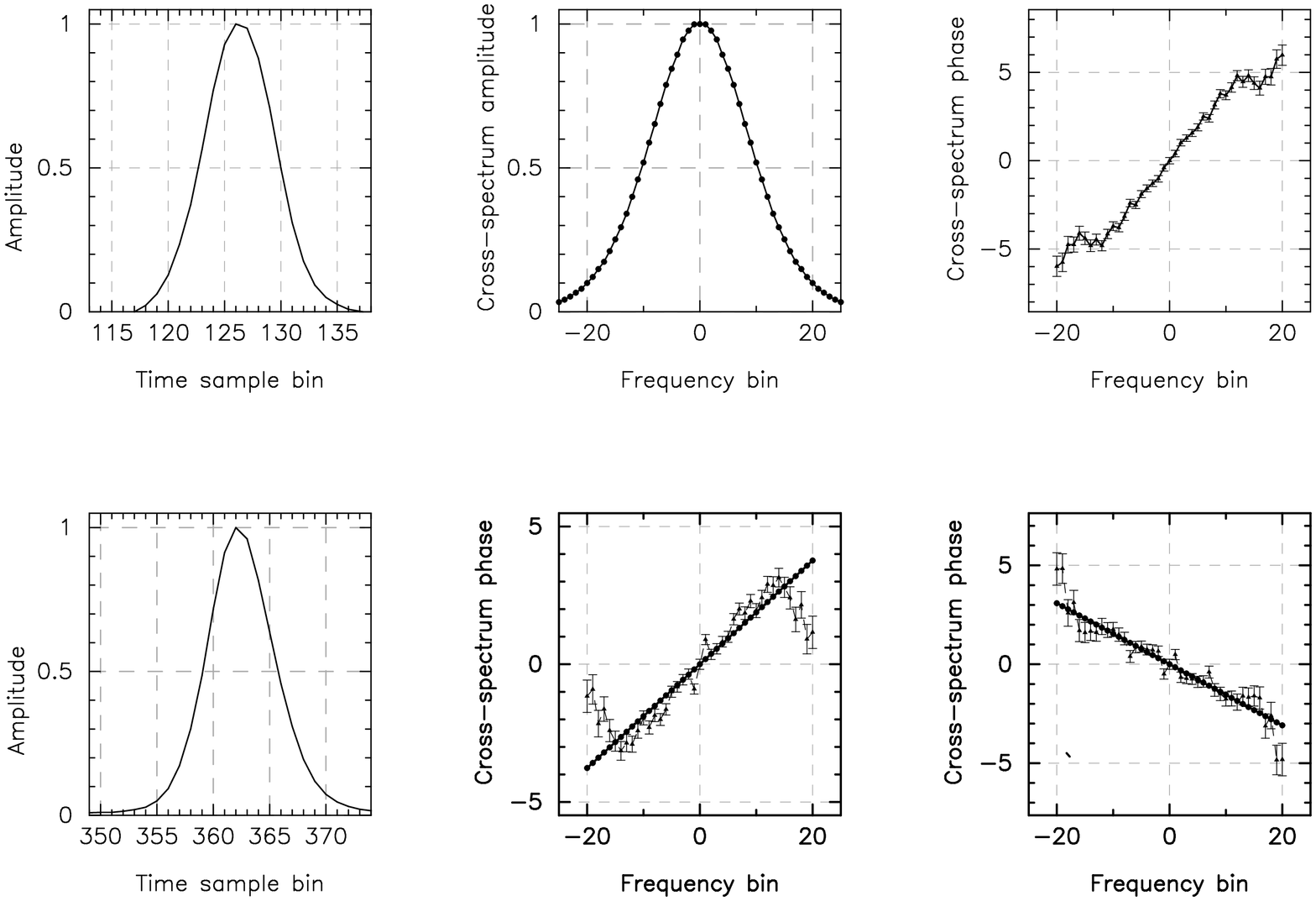}
\caption[[Pulse profiles and CS phase function for PSR B1642$-$03 observed at 610$+$325 MHZ]
{Pulse profiles for PSR B1642$-$03, observed simultaneously at 610 (left upper panel) 
and 325 MHz (left lower panel) bands. Corresponding CS amplitude (middle upper panel), phase function
before template subtraction with best fit line (middle lower panel), CS phase for zero DM fitted 
template (right upper panel) and CS phase function with best fit after template subtraction (right lower panel).}
\label{fig:b1642_profcs_610325}
\end{center}
\end{figure*}
%%%%%%%%%%%%%%%%%%%%%%%%%%%%%%%%%%%%%%%%%%%%% 
%
%%%%%%%%%%%%%%%%%%%%%%%%%%%%%%%%%%%%%%%%%%%%%
\begin{figure*}
\begin{center}
\includegraphics[angle=-90, width=0.6\textwidth]{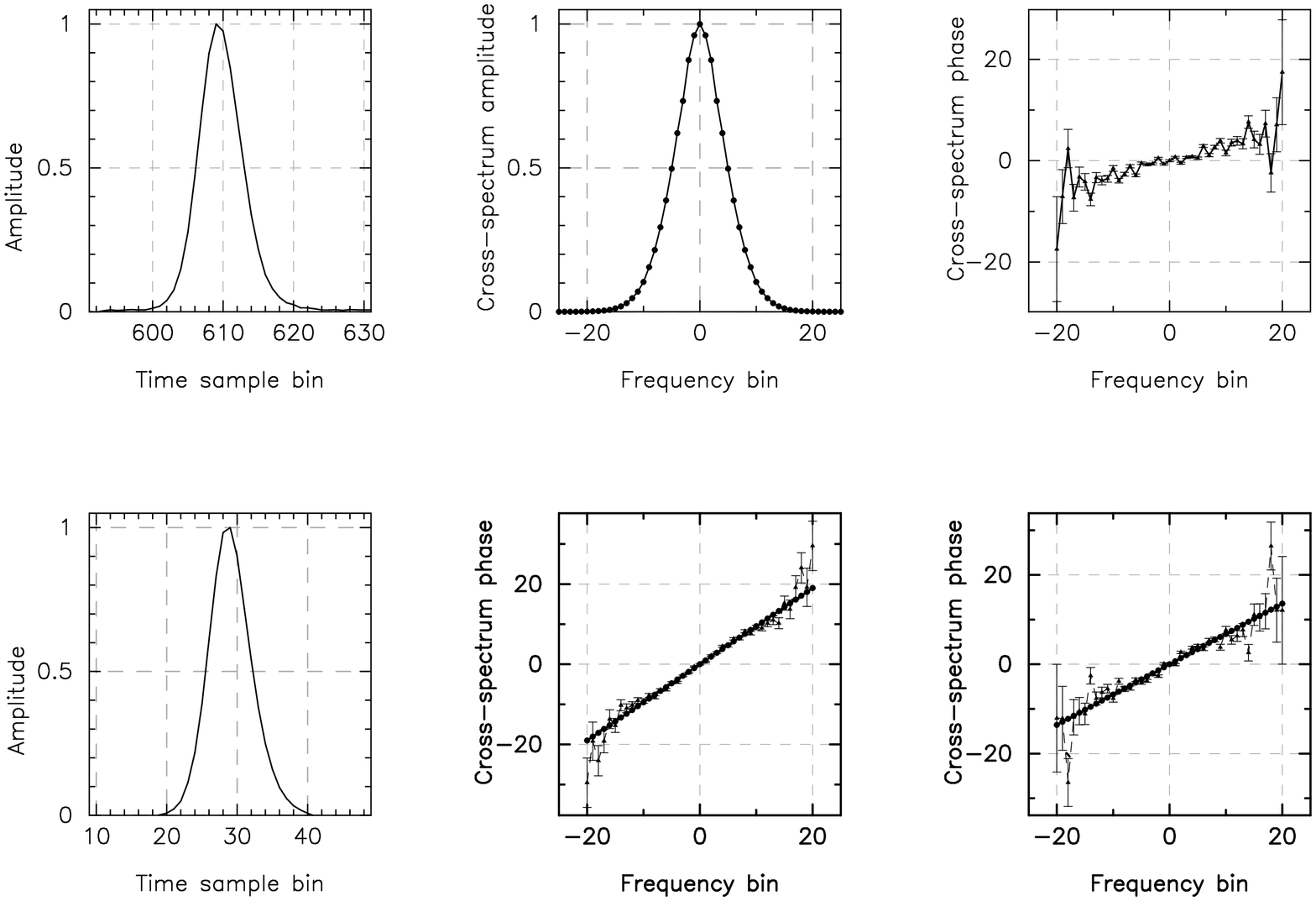}
\caption[[Pulse profiles and CS phase function for PSR B1642$-$03 observed at 325$+$243 MHZ]
{Pulse profiles for PSR B1642$-$03, observed simultaneously at 325 (left upper panel) 
$+$ 243 MHz (left lower panel) bands. Corresponding CS amplitude (middle upper panel), phase function
before template subtraction with best fit line (middle lower panel), CS phase for zero DM fitted 
template (right upper panel) and CS phase function with best fit after template subtraction (right lower panel).}
\label{fig:b1642_profcs_325243}
\end{center}
\end{figure*}
%%%%%%%%%%%%%%%%%%%%%%%%%%%%%%%%%%%%%%%%%%%%% 
%
%%%%%%%%%%%%%%%%%%%%%%%%%%%%%%%%%%%%%%%%%%%%
\begin{figure*}
\begin{center}
\includegraphics[angle=-90, width=0.6\textwidth]{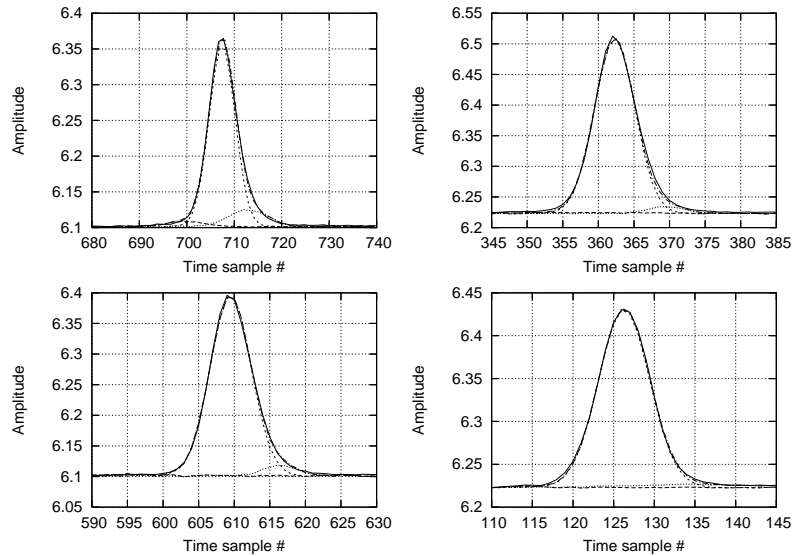}
\caption[PSR B1642_dual_1_2]{Real and fitted pulse profiles, with all the fitting components, for PSR 
B1642$-$03, observed at two frequency bands pairs (simultaneously at each):  325 (left lower panel) $+$ 
243 MHz (left upper panel), and 610 (right lower panel) $+$ 325 MHz (right upper panel) bands.}
\label{fig:b1642}
\end{center}
\end{figure*}
%%%%%%%%%%%%%%%%%%%%%%%%%%%%%%%%%%%%%%%%%%%%%
%
%%%%%%%%%%%%%%%%%%%%%%%%%%%%%%%%%%%%%%%%%%%%%
\begin{figure*}
\begin{center}
\includegraphics[angle=-90, width=0.6\textwidth]{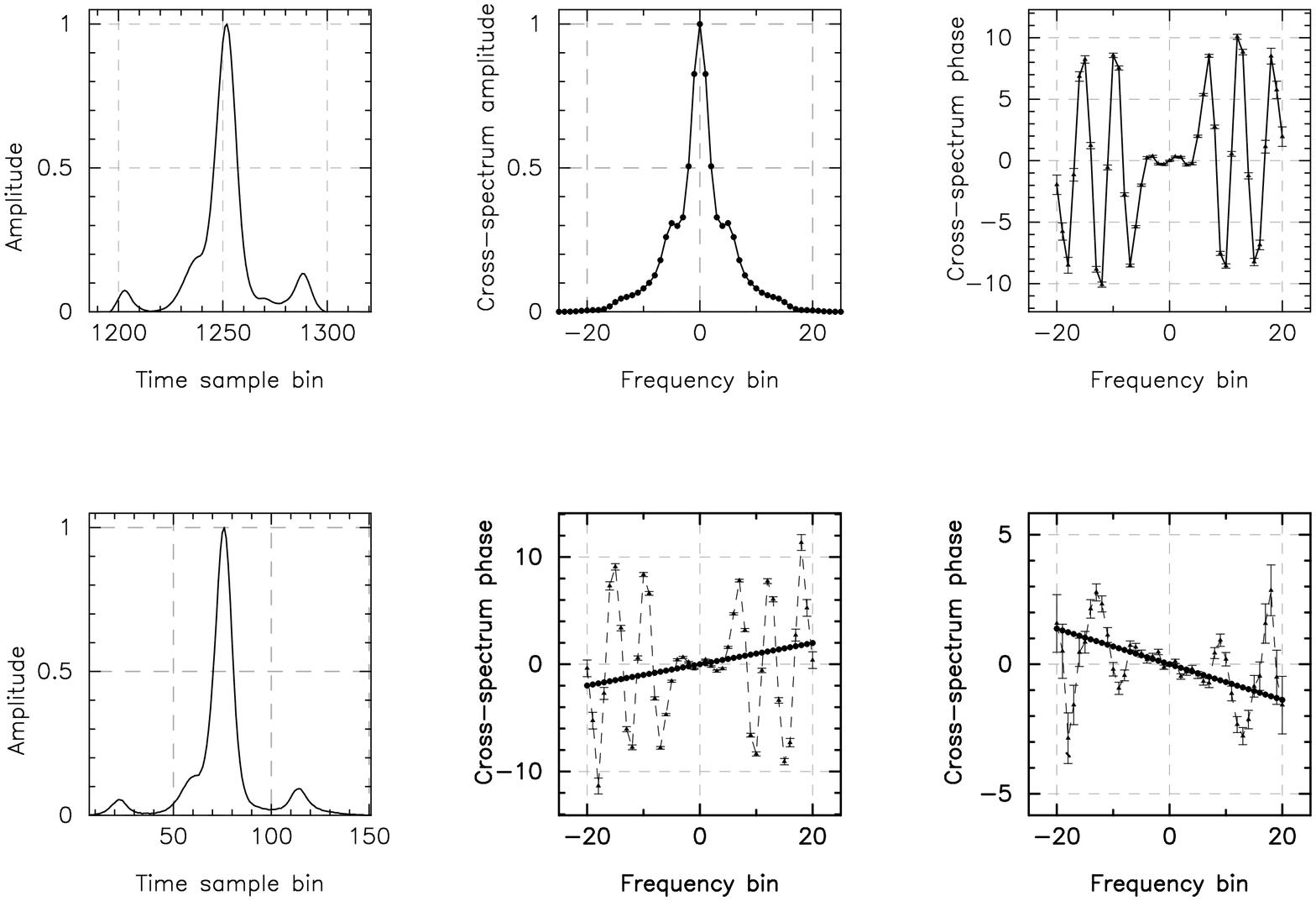}
\caption[[Pulse profiles and CS phase function for PSR B0329$+$54 observed at 610$+$320 MHZ]
{Pulse profiles for PSR B0329$+$54, observed simultaneously at 610 MHz (left upper panel),
320 MHz (left lower panel), $\&$ 243 MHz bands. 
Corresponding to 610$+$320 MHz CS amplitude (middle upper panel), phase function
before template subtraction with best fit line (middle lower panel), CS phase for zero DM fitted 
template (right upper panel) and  CS phase function with best fit, after template subtraction (right lower panel).}
\label{fig:b0329_profcs_610320}
\end{center}
\end{figure*}
%%%%%%%%%%%%%%%%%%%%%%%%%%%%%%%%%%%%%%%%%%%%% 
%
%%%%%%%%%%%%%%%%%%%%%%%%%%%%%%%%%%%%%%%%%%%%%
\begin{figure*}
\begin{center}
\includegraphics[angle=-90, width=0.6\textwidth]{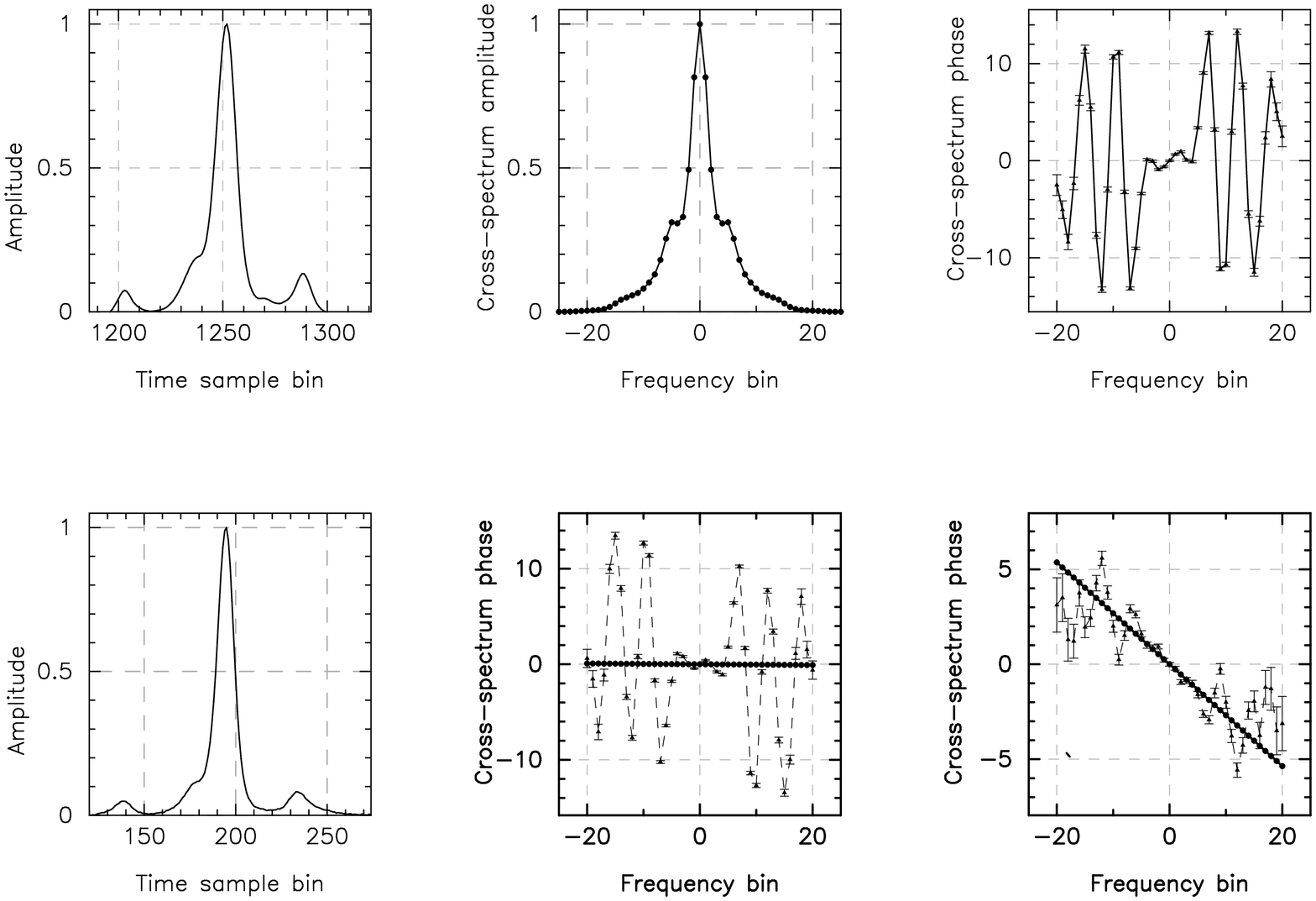}
\caption[[Pulse profiles and CS phase function for PSR B0329$+$54 observed at 610$+$243 MHZ]
{Pulse profiles for PSR B0329$+$54, observed simultaneously at 610 (left upper panel),
320 $\&$ 243 MHz (left lower panel) bands. 
Corresponding to 610$+$243 MHz CS amplitude (middle upper panel), phase function
before template subtraction with best fit line (middle lower panel), CS phase for zero DM fitted 
template (right upper panel) and CS phase function with best fit, after template subtraction (right lower panel).}
\label{fig:b0329_profcs_610243}
\end{center}
\end{figure*}
%%%%%%%%%%%%%%%%%%%%%%%%%%%%%%%%%%%%%%%%%%%%% 
%
%%%%%%%%%%%%%%%%%%%%%%%%%%%%%%%%%%%%%%%%%%%%%
\begin{figure*}
\begin{center}
\includegraphics[angle=-90, width=0.6\textwidth]{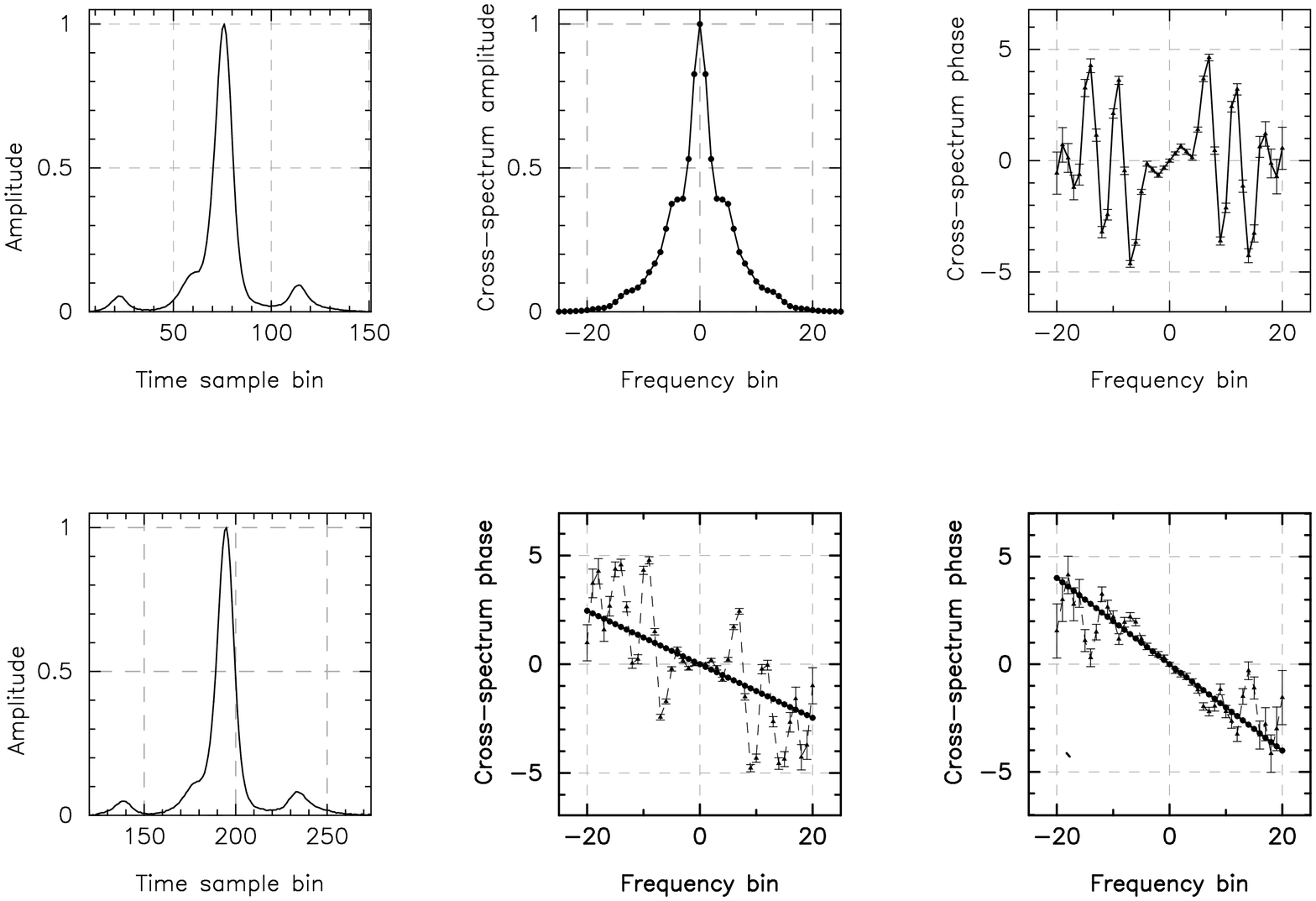}
\caption[[Pulse profiles and CS phase function for PSR B0329$+$54 observed at 320$+$243 MHZ]
{Pulse profiles for PSR B0329$+$54, observed simultaneously at 610 , 320 (left upper panel) 
$\&$ 243 MHz (left lower panel) bands. Corresponding to 320$+$243 MHz 
observation CS amplitude (middle upper panel), phase function
before template subtraction with best fit line (middle lower panel), CS phase for zero DM fitted 
template (right upper panel) and CS phase function with best fit, after template subtraction (right lower panel).}
\label{fig:b0329_profcs_320243}
\end{center}
\end{figure*}
%%%%%%%%%%%%%%%%%%%%%%%%%%%%%%%%%%%%%%%%%%%%% 
%
%%%%%%%%%%%%%%%%%%%%%%%%%%%%%%%%%%%%%%%%%%%%
\begin{figure*}
\begin{center}
\includegraphics[angle=-90, width=0.6\textwidth]{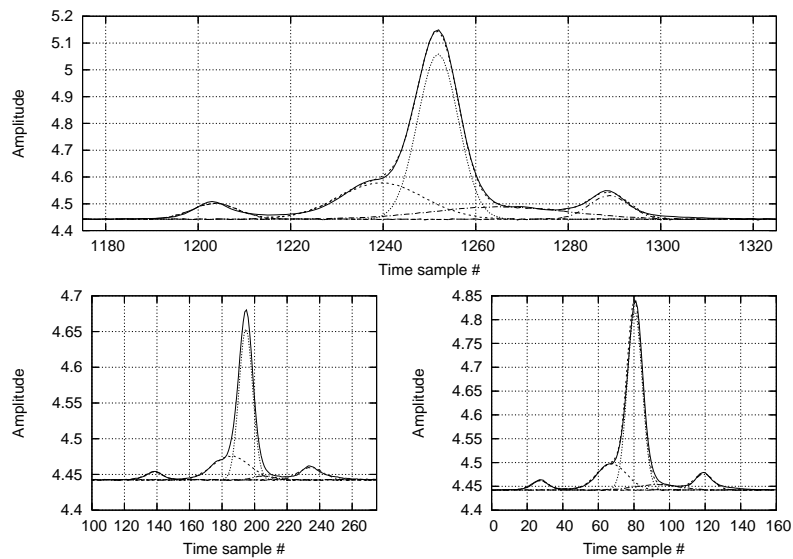}
\caption[PSR B0329$+$54]{The above figure shows pulse profiles for PSR B0329$+$54, observed on January 22, 2001, simultaneously 
at three frequency bands 243 (left lower panel), 320 (right lower panel) and 610(upper panel). The dashed
lines show all the fitted components and the final fitted profile.}
\label{fig:b0329}
\end{center}
\end{figure*}
%%%%%%%%%%%%%%%%%%%%%%%%%%%%%%%%%%%%%%%%%%%%%
%
%%%%%%%%%%%%%%%%%%%%%%%%%%%%%%%%%%%%%%%%%%%%%
\begin{figure*}
\begin{center}
\includegraphics[angle=-90, width=0.6\textwidth]{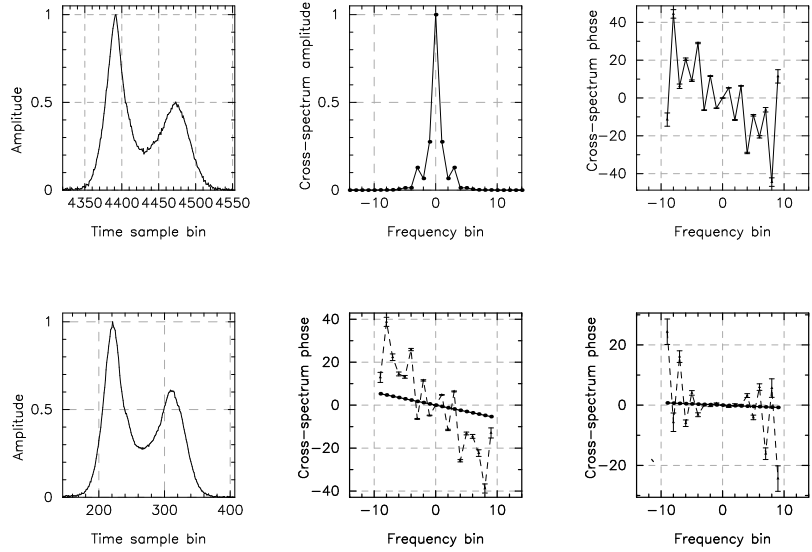}
\caption[[Pulse profiles and CS phase function for PSR B1133$+$16 at 610$+$325 MHZ]
{Pulse profiles and intermediate results for PSR B1133$+$16 observed simultaneously 
at four frequencies. Pulse profiles at 610 (left upper panel) 
$+$ 325 MHz (left lower panel), corresponding CS amplitude (middle upper panel), phase function
before template subtraction with best fit line (middle lower panel), CS phase for zero DM fitted 
template (right upper panel) and CS phase function with best fit, after template subtraction (right lower panel).}
\label{fig:b1133_profcs_610325}
\end{center}
\end{figure*}
%%%%%%%%%%%%%%%%%%%%%%%%%%%%%%%%%%%%%%%%%%%%% 
%
%%%%%%%%%%%%%%%%%%%%%%%%%%%%%%%%%%%%%%%%%%%%%
\begin{figure*}
\begin{center}
\includegraphics[angle=-90, width=0.6\textwidth]{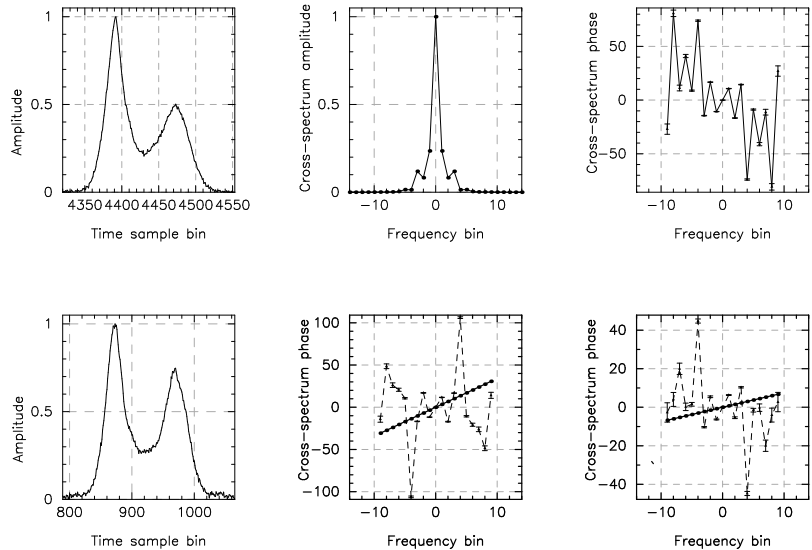}
\caption[[Pulse profiles and CS phase function for PSR B1133$+$16 at 610$+$227 MHZ]
{PSR B1133$+$16 observed simultaneously at four frequencies. 
Pulse profiles at 610 (left upper panel) 
$+$ 227 MHz (left lower panel), corresponding CS amplitude (middle upper panel), phase function
before template subtraction with best fit line (middle lower panel), CS phase for zero DM fitted 
template (right upper panel) and CS phase function with best fit, after template subtraction (right lower panel).}
\label{fig:b1133_profcs_610227}
\end{center}
\end{figure*}
%%%%%%%%%%%%%%%%%%%%%%%%%%%%%%%%%%%%%%%%%%%%% 
%
%%%%%%%%%%%%%%%%%%%%%%%%%%%%%%%%%%%%%%%%%%%%%
\begin{figure*}
\begin{center}
\includegraphics[angle=-90, width=0.6\textwidth]{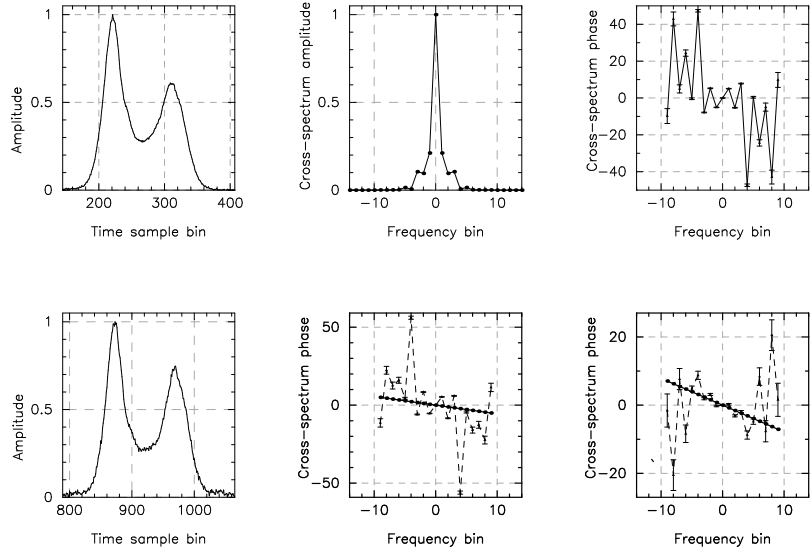}
\caption[[Pulse profiles and CS phase function for PSR B1133$+$16 at 325$+$227 MHZ]
{ Pulse profiles at 325 (left upper panel) $+$ 227 MHz (left lower panel) for PSR B1133$+$16, 
observed simultaneously at four frequencies. Corresponding CS amplitude (middle upper panel), phase function
before template subtraction with best fit line (middle lower panel), CS phase for zero DM fitted 
template (right upper panel) and CS phase function with best fit, after template subtraction (right lower panel).}
\label{fig:b1133_profcs_325227}
\end{center}
\end{figure*}
%%%%%%%%%%%%%%%%%%%%%%%%%%%%%%%%%%%%%%%%%%%%% 
%
%%%%%%%%%%%%%%%%%%%%%%%%%%%%%%%%%%%%%%%%%%%%
\begin{figure*}
\begin{center}
\includegraphics[angle=-90, width=0.6\textwidth]{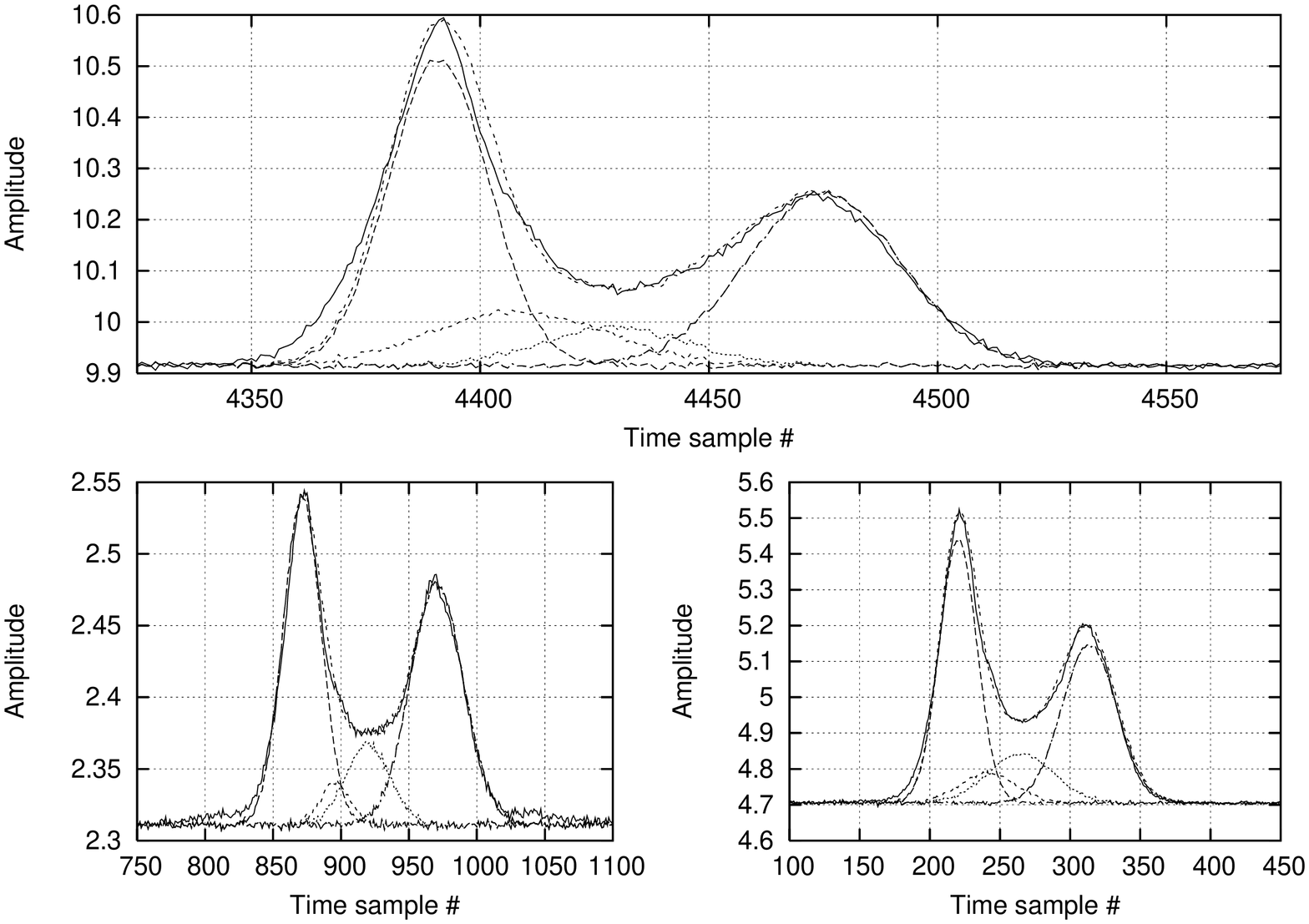}
\caption[PSR B1133$+$16]{The above plot shows pulse profiles for PSR B1133$+$16 observed on February 3, 2004, 
simultaneously at three frequency bands 227 (lower left panel), 325 (lower right panel) and 610 MHz 
(upper panel). The dashed line shows
all the fitted components and the final fitted profile at the three frequency bands.}
\label{fig:b1133}
\end{center}
\end{figure*}
%%%%%%%%%%%%%%%%%%%%%%%%%%%%%%%%%%%%%%%%%%%%%
%
%%%%%%%%%%%%%%%%%%%%%%%%%%%%%%%%%%%%%%%%%%%%%%%%%%
%
\section {Frequency dependent DM variations} \label{sec:res}
%%%%%%%%%%%%%%%%%%%%%%%%%%%%%%%%%%%%%%%%%%%%%%%%%%
%[some of this first para may already be covered in section 2 ??]
Tables \ref{tab:b1642}, \ref{tab:b0329} and \ref{tab:b1133} give the DM values 
obtained for PSR B1642$-$03, PSR B0329$+$54 
and PSR B1133+16 respectively.
Column 1 of these tables (\ref{tab:b1642}, \ref{tab:b0329} and \ref{tab:b1133}) 
represent the frequency pair used for the DM analysis. Column 2 of the tables 
provides the information about the DM per bin delay between signals at the 
two frequency bands. Column 3 gives the DM obtained from the various frequency pairs,
and clearly significant DM changes with frequency are seen. In this section we primarily 
address the question if pulse shape evolution can cause such DM changes.

{\bf PSR B1642-03:} This pulsar has a relatively simple pulse profile and is seen to 
have a single emission component across frequencies. The lower middle panels of 
figures~\ref{fig:b1642_profcs_610325} and \ref{fig:b1642_profcs_325243} show the phase 
function of the CS for frequency pairs 610$+$325 MHz and 325$+$243 MHz, respectively. 
The phase function for 610$+$325 MHz is seen to be linear close to the zero frequency
bin, with a sharp bend towards the edge frequencies. A similar trend but with much 
lower prominence for the sharp bend is seen for the frequency pair 325$+$243 MHz.
To assess if the phase function has any contribution from pulse shape evolution, we 
have fitted Gaussian components to the profiles and found that the profile is described 
adequately, at each frequency, by three Gaussians -- a central dominating Gaussian and 
two weaker Gaussian on either side, as shown in figure~\ref{fig:b1642}.  
Assuming the peak of the central component to be the fiducial point, we find the zero 
DM phase functions of the CS to be as shown in the top right panel of 
figures~\ref{fig:b1642_profcs_610325} and ~\ref{fig:b1642_profcs_325243}. Considering 
these as the template phase functions due to pulse shape evolution, we obtain the 
template subtracted phase functions as shown in the bottom rightmost panel of these
figures. Clearly, we can recover the linear gradient of the phase function which is 
primarily due to $\Delta t_f$. The template subtracted DMs are given in column 5 of 
Table \ref{tab:b1642}, where we still see a significant change of DM between the two
frequency pairs.  In analogy with our simulations, since the profile evolution for 
PSR B1642$-$03 is relatively symmetric, we find that the template subtracted DM is quite 
similar to the non template subtracted DM. 

For PSR B1642$-$03 \cite{Shitov_etal} reported frequency dependent variation of DM, where
the DM observed from low frequency pairs (60 and 102.5 MHz) yielded a value of 35.736(5)
$\rm{pc ~cm^{-3}}$ while higher frequency data (408, 610, 925, 1420 MHz) by \cite{Hun_1971} 
gave DM=35.665(5) $\rm{pc ~cm^{-3}}$.  Our DM estimates, through restricted over a narrower 
frequency range, show the opposite trend i.e low frequency pair (325$+$243 MHz) showing 
larger DM compared to the value from the higher frequency pair (610$+$325 MHz). The 
magnitude of the DM variation is only a factor of 2.2 less than that estimated by 
\cite{Shitov_etal}. Since the data used by \cite{Shitov_etal} is spaced over 17 years 
and can be influenced by the temporal DM changes in the ISM, we believe that their 
conclusion about the observed DM variation is less robust. 

{\bf PSR B0329$+$54:} This is a multicomponent pulsar with a strong central component
and weak outriders.  As seen in figures~\ref{fig:b0329_profcs_610320}, 
\ref{fig:b0329_profcs_610243} and \ref{fig:b0329_profcs_320243}, the phase function 
of the CS shows a significant oscillatory behaviour riding on a linear gradient. The 
profile is well fitted by five Gaussians, at all the frequencies, as shown in 
figure~\ref{fig:b0329}. Using the peak of the central component as the fiducial point, 
and performing a similar exercise of finding the zero DM phase template and subtracting 
it from the actual phase of the CS, we find that the resultant phase shows the distinct 
linear gradient, as illustrated in the rightmost bottom panels of 
figures~\ref{fig:b0329_profcs_610320}, \ref{fig:b0329_profcs_610243} and 
\ref{fig:b0329_profcs_320243}. There is some residual oscillatory behaviour left at a
lower level towards the edge frequency bins, which could be attributed to our 
inadequacy to fit the profile exactly with Gaussian components. In Table~\ref{tab:b0329}, 
the DM values quoted in columns 3 and 5 before and after template subtraction, respectively,
however are very similar and show a significant dependence with frequency. The similarity
of the DM values are possibly due to the fact that PSR B0329$+$54 has a very strong central
component which dominates any other structure in the pulse. Hence the phase function of 
the CS close to the central frequency bins (as seen in the bottom middle panels of 
figure~\ref{fig:b0329_profcs_610320}, \ref{fig:b0329_profcs_610243} and 
\ref{fig:b0329_profcs_320243}) has the characteristic linear gradient with small errors 
and hence the straight line fit to the phase responds primarily to this linear gradient. 
Alternatively, since the central component is a dominating component and we think that 
the fiducial point lies at the peak of the component, we can put a window only on the 
central component and use only the windowed region of the pulse to estimate the DM. In 
this method the corruption of the phase function of the CS will be significantly reduced, 
since now we are dealing with only a single component which evolves more or less 
symmetrically with frequency. We have applied this technique to the data and
found the phase function of the CS to become more linear, however the corresponding DMs
found were comparable to values quoted in column 5 of Table~\ref{tab:b0329}. 
For PSR B0329$+$54 we find a similar trend like PSR B1642$-$03, where the high frequency 
pairs give larger DMs compared to low frequency pairs.  For PSR B0329$+$54, as far as we 
know, no such DM variation study has been reported earlier.

{\bf PSR B1133$+$16:} The profile of this pulsar shows a great deal of asymmetry, with 
two prominent conal components of different intensity level, present along with a central 
bridge of emission.  The conal components vary significantly with frequency also, and 
this results in a complicated phase function of the CS, similar to what we have encountered 
in our simulation experiment number 6 earlier (see Table \ref{tab:simul_res}). The DM 
estimates between frequency pairs vary significantly, with estimated fractional time sample 
bin delays exceeding a few integer bins (for e.g. 4.985 bin for the frequency combination 
610$+$227 MHz, as seen in column 4 of Table~\ref{tab:b1133}). Such large changes in DM 
are clearly due to the complicated phase function of the CS as seen in the bottom middle 
panels of figures~\ref{fig:b1133_profcs_610325}, \ref{fig:b1133_profcs_610227} and 
\ref{fig:b1133_profcs_325227}. The phase function in each case has a significant amount 
of ringing or oscillatory behaviour, and attempts to fit straight lines to the data are 
clearly unsatisfactory.  

To understand this large discrepancy we again attempt to estimate the intrinsic phase 
function of the CS due to pulse shape variations.  The pulse profiles at each frequency 
is adequately fitted by five Gaussians (see figure ~\ref{fig:b1133}).  Since there is no 
clear central core component in this pulsar, the choice of the fiducial point to produce 
the template phase function by aligning the profiles at zero DM is tricky.
However by choosing the peak of the centrally placed Gaussian (third Gaussian from left), 
we obtain the zero DM template and the template subtracted DM as shown in the 
top and bottom right panels of figures~\ref{fig:b1133_profcs_610325},
\ref{fig:b1133_profcs_610227} and \ref{fig:b1133_profcs_325227}, respectively.
The DM differences obtained based on fitting the template subtracted phase functions are 
far less varying with frequency as seen in columns 5 and 6 of Table~\ref{tab:b1133})
(less than an integer bin). It is indeed remarkable that by choosing the central point
in the profile we are able to get reasonable DM values between frequencies. However, since 
the resultant template subtracted phase functions are not entirely linear, we are not 
certain if the finer DM difference seen in column 5 of the table are real or still 
corrupted by the intrinsic phase function. We have also tried by choosing a few other 
points in the profile as the fiducial points, but do not find any better convergence in 
the DM values with frequency pairs.

PSR B1133$+$16 has been a subject of several multi-frequency DM studies. 
Based on alignment of micro-structure in the pulse \cite{Bor_1983} and \cite{Hankins-91} 
obtained a DM = 4.8460(1) $\rm pc ~cm^{-3}$ (from frequencies 318 and 111 MHz) 
while \cite{Pop_etal_1987} obtained DM=4.8413(1) $\rm pc ~cm^{-3}$ (using three frequencies 
taken pairwise between 103, 79 and 68 MHz). By alignment of multi-frequency average profile 
data using the midpoint of the outer conal pair as a fiducial marker \cite{Hankins-91} found
DM=4.8470(3) $\rm pc ~cm^{-3}$ (based on frequency ranging from 24.8 MHz to 4870 MHz)
and \cite{PnW-92} observed DM=4.8471(2) $\rm pc ~cm^{-3}$, where the 
observed DMs were slightly larger than that obtained from the micro-structure methods.
The DM values reported by us in Table~\ref{tab:b1133} are larger than the previously quoted 
value, which could be a result of temporal variation of DM in the ISM, and is significantly
larger than the micro-structure based DM. The old and new catalog value quoted in 
Table~\ref{tab:pul_param} also shows a significant variation and supports this hypothesis 
of temporal variation. \cite{Shitov_etal} claimed that the low frequency pulses showed higher 
DM values compared to high frequencies, which however was later disputed by \cite{Phi_1991} 
who found that the pulse arrival times were consistent with the cold plasma dispersion law. 
Because of the complexity in finding finer changes in DMs as mentioned earlier, and a smaller 
frequency coverage of our data set, we are unable to comment any more conclusively on this 
matter, based on our analysis.
 
%
%%%%%%%%%%%%%%%%%%%%%%%%%%%%%%%%%%%%%%%%%%%%%%%%%%%%%%
\section{Discussion and Conclusions} 		\label{sec:con}
%%%%%%%%%%%%%%%%%%%%%%%%%%%%%%%%%%%%%%%%%%%%%%%%%%%%
%
Earlier studies regarding the existence of frequency dependent DM variations in pulsars 
have resulted in very different conclusions. Based on multi-frequency data and pulsar 
timing analysis, \cite{Cra_1970} and \cite{PnW-92} have shown that pulsar DMs obey the 
cold plasma dispersion law given by equation~\ref{eq1}. On the other hand \cite{Shi_Mal_1985}, 
\cite{Kuz_1986} and \cite{Shitov_etal} claim that DMs at low frequencies are systematically 
larger than DMs at high frequencies. This phenomenon usually goes by the name `super 
dispersion', and was explained as magnetic field sweep back in the pulsar magnetosphere.
\cite{PnW-92} have claimed departure from equation~\ref{eq1} at high frequencies, for two 
pulsars.  As pointed out by \cite{Phi_1991}, the comparison of high and low frequency DMs 
is often based on non simultaneous data  and hence the DMs obtained at different frequencies 
may differ due to temporal DM variability caused by the ISM. Another plausible cause for DM 
changes claimed by different observers might be due to the choice of fiducial point, used 
in the analysis to find the DM, not being uniform or consistent.

Results presented in this paper and {\it Paper I} are based on simultaneous observations 
of the frequency pairs for PSR B1642$-$03 (where the two frequency pairs are observed within 
typically half an hour interval), and three and four frequency simultaneous observation for 
PSR B0329$+$54 and B1133$+$16.  Hence, these are not affected by temporal changes in the ISM.
Our DM analysis technique and high sensitivity of the GMRT allows us to obtain DMs, with 
an accuracy of ~1 part in $10^4$, which is either comparable or better than accuracies 
obtained from previous studies.  In this paper we have considered the possibility that 
pulse shape evolution with frequency can affect DM estimates, resulting in pseudo 
DM variations.  We have devised a zero DM template subtraction method to eliminate this 
effect and still found that DM for our sample pulsars varies with frequency, to some extent.  
This variation is seen for all the epochs of observations for PSR B1642$-$03 (for e.g. see 
Figure 5 in {\it Paper I} which shows the DM variation results for B1642$-$03 for several 
epochs) and PSR B0329$+$54.  For both these pulsars the DMs obtained from low frequency pairs 
are less than the higher frequency pairs. In both these cases we have chosen the peak of the 
pulse profile (by a Gaussian fitting procedure) as the fiducial marker.  In the case of 
PSR B1133$+$16 we found that pulse shape evolution can affect DM estimates by significant 
amount. While our zero DM template subtraction method could produce gross correction of 
the DM values, resulting in matching DMs between frequency pairs at a level of in 1 part 
in $10^{-2}$, the differences are still significant w.r.t the errors in estimation.
However our subtraction method is not accurate enough to claim that these variations
are real.

We believe that we have established with certainty that PSR B1642$-$03 and PSR B0329$+$54 
show DM variation with frequency where high frequency DM pairs show higher DM than the low 
frequency pairs. This effect is opposite to `super dispersion' claimed by several authors 
\citep[e.g.][]{Shitov_etal}.  Physically, DM variation with frequency can result both due 
to the ISM \citep[e.g.][]{Ram_etal_2006} and the pulsar magnetosphere.  For a uniform medium 
in the ISM, dispersion depends on the column density of electrons along the line of sight from 
the source to the observer. However for a turbulent ISM, due to refractive effects, the 
emission from the source performs multi-path propagation causing the effective DM to vary. 
Since pulsar beams at lower frequencies are larger than at higher frequencies, the geometry 
for multi-path propagation and distribution of the turbulent features becomes a function of 
frequency, resulting in DM variation with frequency. In such a case, the sign of the DM change 
with frequency is unpredictable.  Another related ISM effect which can cause DM variation is 
connected to the change of the pulsar waveform due to scatter broadening. Scattering causes 
the intrinsic emission from the pulsar to be convolved with the response function (typically 
observed to have an exponential form) of the ISM. Since scattering is a strong function of 
frequency $\nu^{-4.4}$ \citep[see][and references therein]{Ram_etal_1997}, the fiducial 
marker in the pulse profile, tends to arrive later at lower frequencies. This effect will 
cause DMs to be larger at lower frequencies, quite opposite to what we observe, and hence 
can be ruled out as a possible explanation for DM variation for PSR B1642$-$03 and B0329$+$54.

Alternatively, the emission process and propagation effects in the pulsar magnetosphere
can give rise to frequency dependent DM variations. Our observations suggest that the 
fiducial markers tend to arrive earlier at lower frequencies. For both PSR B1642$-$03 and 
B0329$+$54 the fiducial point is chosen to be the peak of the central core emission which 
is thought to lie in the plane containing the rotation axis and the magnetic axis. If we 
assume that RFM operates for core emission in pulsars, then progressively lower frequencies 
are emitted from higher heights above the neutron star surface. For such an emission process, 
aberration can cause the low frequency emission to arise at earlier rotation phase compared 
to the higher frequencies.  Such an effect can give rise to DM dependent frequency variations. 
Combination of RFM and presence of dispersive plasma in the pulsar magnetosphere can introduce 
differential shifts in the pulse profile as a function of frequency. 

Currently more observations are needed both over a wider frequency range and for a number 
of pulsars to constrain the nature of frequency dependent DM variations in pulsars. 
Also further progress in pulsar emission and propagation theories are needed to
understand these effects.

\vspace{5mm}
\noindent{\large\bf Acknowledgments:}
We would like to thank Ajit Kembhavi for his co-operation. We would also like to thank 
M. Kramer for providing us his Gaussian fitting program which is used extensively in this 
work.  We would like to thank the staff at the GMRT for making the observations possible.
The GMRT is run by the National Centre for Radio Astrophysics.

\label{lastpage}

\end{document}